# Thalamocortical interactions shape hierarchical neural variability during stimulus perception

Adrià Tauste Campo[1,*], Antonio Zainos[2], Yuriria Vázquez[2,a], Raul Adell Segarra[1], Manuel Álvarez[2], Gustavo Deco[3,4], Héctor Díaz[2], Sergio Parra[2], Ranulfo Romo[2,5,b,*], Román Rossi-Pool[2,6,*]

[1]Computational Biology and Complex Systems group, Department of Physics, Universitat Politècnica de Catalunya, Avinguda Dr. Marañón, 44-50, 08028 Barcelona, Catalonia, Spain

[2]Instituto de Fisiología Celular─Neurociencias, Universidad Nacional Autónoma de México, 04510 Mexico City, Mexico

[3]Center for Brain and Cognition (CBC), Department of Information Technologies and Communications (DTIC), Pompeu Fabra University, Edifici Mercè Rodoreda, Carrer Trias I Fargas 25-27, 08005, Barcelona, Catalonia, Spain

[4]Institució Institució Catalana de Recerca I Estudis Avançats (ICREA), Passeig Lluis Companys 23, 08010, Barcelona, Catalonia, Spain

[5]El Colegio Nacional, 06020 Mexico City, Mexico

[6]Centro de Ciencias de la Complejidad, Universidad Nacional Autónoma de México, Mexico City, Mexico.

[a]Present address: Laboratory of Sensory Neuroscience, The Rockefeller University, New York, NY 10065.

[b]Present address: El Colegio Nacional, 06020 Mexico City, Mexico

*Correspondence: adria.tauste@upc.edu (A.T.C.); ranulfo.romo@gmail.com (R.R.); romanr@ifc.unam.mx (R.R.-P.)

**The brain is hierarchically organized to process sensory signals. But, to what extent do functional connections within and across areas shape this hierarchical order? We addressed this problem in the thalamocortical network, while monkeys judged the presence or absence of a vibrotactile stimulus. We quantified the variability by means of intrinsic timescales and Fano factor, and functional connectivity by means of a directionality measure in simultaneously recorded neurons sharing the same cutaneous receptive field from the somatosensory thalamus (VPL) and areas 3b and 1 from the somatosensory cortex. During the pre-stimulus periods, VPL and area 3b exhibited similarly fast dynamics while area 1 showed much slower timescales. Furthermore, during the stimulus presence, the Fano factor increased along the network VPL-3b-1. In parallel, VPL established two separate main feedforward pathways with areas 3b and 1 to process stimulus information. While feedforward interactions from VPL and area 3b were favored by neurons within specific Fano factor ranges, neural variability in area 1 was invariant to the incoming pathways. In contrast to VPL and area 3b, during the stimulus arrival, area 1 showed significant intra-area interactions, which mainly pointed to neurons with slow intrinsic timescales. Overall, our results suggest that the lower variability of VPL and area 3b regulates feedforward thalamocortical communication, while the higher variability of area 1 supports intra-cortical interactions during sensory processing. These results provide evidence of a hierarchical order along the thalamocortical network.**

## INTRODUCTION

Perception results from the interaction of neural networks that follow a hierarchical connectivity organization that extends from the receptors up to the cortex (1–3). The thalamus is known to play a key role in gating the flow of information to the cortex (4). To unveil the intricacies of the thalamocortical network, several past studies have investigated the feedforward and feedback connections that link the thalamus and cortex (5–7). On this matter, a vast number of experimental studies on different sensory modalities (8–10) have provided a detailed model of the flow of sensory information processing in the thalamocortical network. This evidence has shown that it is composed of first order thalamic nuclei that act as sensory relay (feedback-driven) between the brainstem and the cortex, and of higher order nuclei that mediate cortico-cortical communications (4, 11, 12). In either case, thalamic nuclei have been shown to be drivers of cortical signals in contexts such as sensory representation (13, 14) and sensory adaptation (15, 16).

Over the last decades, the structural picture of the thalamocortical network has been refined thanks to functional connectivity studies on electrophysiological simultaneous recordings in both anesthetized (17) and awake animals (18, 19). Moreover, we have recently described the level of feedforward and feedback single-neuron functional interactions between the ventral posterior lateral (VPL) nucleus of the thalamus and areas 3b and 1 of the primary somatosensory (S1) cortex during a vibrotactile detection task performed by trained monkeys (20). The results showed that feedforward prevailed over feedback interactions during stimulus perception and indicated the presence of zero-lag synchronization as a putative hallmark of active thalamocortical transmission. However, how these interactions were integrated into the hierarchical connectivity organization of the somatosensory network (21–24) remained unaddressed. In particular, it is unclear how functional connections in the thalamocortical network are constrained by the local properties of VPL (25, 26) and subareas 3b and 1 of S1 (27, 28). For instance, how are somatosensory interactions related to the capacity of neurons in each area to encode information? Are the interactions between VPL and each subarea of S1 different enough to support the relay role of area 3b? Or is VPL interacting analogously with neurons from areas 3b and 1? Further, are the intra- and inter-area interactions affected by the response variability of the neurons (29–31)? Recently, the study of the brain's hierarchical organization through the estimation of neural timescales (or time constants) has become a topic of intense research (21, 32–34). Specifically, the area's intrinsic timescale provides a quantification of the period in which the neurons of an area can integrate their input and is often associated with the strength of intra-area recurrent excitation (1). In this context, it is therefore natural to ask: how are functional interactions in the somatosensory network related to the capacity of the neurons to integrate inputs? In other words, are the intra-area interactions related to the intrinsic timescale of each network?

In the present study, we investigate the above questions by first measuring neural variability across two dimensions (inter-trial and intra-trial) as a proxy to infer the distribution of Fano factors and timescales of each area and assess their concordance with their hierarchical organization (1). Upon unraveling directed interactions across VPL and areas 3b and 1 during the stimulus presence, we report that VPL establishes transient parallel feedforward connections with area 3b and area 1, respectively, that are concurrent to intra-area 1 connectivity. In other words, area 3b->area 1 can be regarded as a secondary feedforward station in the touch processing route. In addition, while outgoing feedforward connections are mainly supported by small thalamic and large area 3b Fano factors, intra-area 1 connectivity is supported by cortical neurons with high timescales. Thus, our refined analysis of thalamocortical functional connectivity (20) is shown to be integrated into the somatosensory hierarchical organization (23) reflecting a gradation along the pathway VPL-area 3b-area 1, by which the individual variability of neurons from VPL and area 3b is particularly tuned to facilitate feedforward communication, while the variability of neurons from area 1 is aimed to supporting recurrent cortical connections. These results provide strong evidence that a hierarchical order in the somatosensory network could be established by employing variability measures, intrinsic timescales and directed information (DI) measures across the different areas of the thalamocortical network.

**RESULTS**

To study the mutual influence between neural interactions and local firing rate variability, we analyzed the thalamocortical neuronal recordings obtained in four trained monkeys during a vibrotactile detection task in which the monkey received a mechanical vibration of variable amplitude and had to report whether the stimulus was present or absent by pressing a push button (Fig 1A, e.g. See (18, 25, 35) for details). Specifically, during those trials in which monkeys correctly performed the task, denominated as stimulus-present hits, or stimulus-absent correct rejections (Fig 1B), we analyzed the time-varying activity of neurons sharing receptive fields from the VPL nucleus (n=96 neurons) and S1 neurons (n=420; see Fig. 1C-D). Here, we classified the neuronal activity of S1 into two subareas: area 1 (n=336) and area 3b (n=84; Fig. 1C-D). Raster plots of nine neurons recorded from the three areas are shown as exemplary cases in Fig. S1. First, we examined the mean firing rate in each population during stimulus-present (> 8 μm) hit trials and correct rejections (CR; Fig 1B). This firing-rate analysis revealed at least three main electrophysiological stages for supra-threshold stimuli (Fig 1E, top), while the responses during CRs maintained a consistent differentiation between the three areas (Fig. 1E, bottom). Specifically, during the first half of the stimulus period (0-0.25s), area 1 showed larger firing rates than the converging values of area 3b and VPL. Note that during the second half of the stimulus (0.25-0.5s), the firing rate of the three areas decreased to a similar average value. The latency distributions in each area (Fig. 1F) highlighted the shift tendency observed across the somatosensory processing: VPL-3b-1.

To unravel the link between neural variability and thalamocortical interactions during stimulus perception, we explored two sources of spiking variability. 1) Within-trial temporal variability, measured by the autocorrelation decay parameter known as the *intrinsic timescale* (32). 2) Inter-trial variability, measured by the *Fano factor*, evaluated across trials recorded during a fixed experimental condition. We next characterize the neuronal activity of each area according to both sources of variability.

**Intrinsic timescales exhibit a thalamocortical hierarchical organization.**

To infer the intrinsic timescales of the recorded populations in VPL, area 3b and area 1, we followed (32) and more recent works (33, 34). The timescale constant studied for this purpose was the exponential decay rate of the basal (resting state or foreperiod) autocorrelation. We focused on the spontaneous spiking activity of each unit during a common fixed period of 1.5s before stimulation onset. We separately estimated the intrinsic timescale for single neurons and for each area (Methods). In both cases, we calculated and fit the corresponding autocorrelation function (either one per neuron or one per area) via an exponential function with $\tau$ parameter equal to the intrinsic timescale. First, to obtain a $\tau$ for each neuron separately, we constructed the autocorrelation function by averaging accumulated autocorrelation values across time bin pairs of the same temporal distance. Then, to estimate densities, we only considered the neurons that showed sufficient convergence guarantees in the parameter fitting solution (Methods). When inspecting the timescale distribution in each area (Fig. 2A), the broader distribution of area 1 in contrast to VPL suggested a larger diversity of autocorrelation decay rates in this sensory cortex than in the thalamus. Furthermore, the intrinsic timescale was significantly smaller in VPL than in area 1 (VPL, median $\tau$ = 8.46ms; area 1, median $\tau$ = 12.15ms, Ranksum test, $P<0.05$). Moreover, neurons in area 3b were positioned in between (median $\tau$ = 8.64ms) showing timescales which were closer to VPL values, and which were significantly smaller than in area 1.

On the other hand, to obtain intrinsic timescales at the population level, we constructed a single autocorrelation function per area by averaging pooled values across neurons and time bin pairs. The parameter-fitting results obtained with this second approach (Fig. 2B-D) were shown to be stable for VPL ($\tau$ = 9.98ms, Fig. 2B) while they were slightly shifted to greater values for area 3b (τ = 10.11ms, Fig 2C) and area 1 (τ = 14.06ms, Fig 2D). Taken together, these results exhibited the same hierarchical ordering than the median timescales obtained from single-neuron distributions, reflecting how the timescale becomes longer across the somatosensory network. In particular, the resulting timescales suggest a larger average level of integration of S1 as compared to the VPL in agreement with previous results obtained in the rodent visual pathway (34). These findings also pinpoint the existence of specific neural dynamics associated with different regions (area 3b, area 1) within S1.

**The Fano factor correlates with the hierarchical somatosensory network.**

To complement the previous analysis, we also examined the inter-trial variability inherent to each area, estimating the Fano factor of each neuron across correct trials (Fig 3, Methods). The Fano factor measures the overdispersion of a given multiple-trial sample with respect to the Poisson distribution under a fixed experimental condition (29–31, 36). Importantly, it has been recently discussed that the practical estimation of the Fano factor can be largely biased by neuronal firing rate fluctuations (31, 37) and its applications on spiking datasets exhibiting super-Poisson variability during stimulus presence is questionable (30). Nevertheless, previous studies have employed this metric to establish a hierarchy across the visual pathway (38), to differentiate different network states (39) or to exhibit a widespread mechanism to decrease variability across the cortex (36).

Here, we estimated the Fano Factor through the slope of the best linear fit of data formed by each pair of mean and standard deviation across each stimulus-amplitude value. By this approach, we obtain a single value per neuron and plot its variation during the time course of stimulus-present trials in each area (Fig 3A). A visual inspection of the dynamic plots in Fig 3A suggested that both the baseline average value and the stimulus-driven decline increased across the hierarchy. Particularly, while area 1 exhibited a much higher Fano factor during the basal period, it diminished much more sharply during the stimulus. Again, the Fano factor values during the basal period and its decline, visually established a hierarchical processing across the somatosensory network.

To further analyze this decline during the stimulus period, we examined the mean-to-variance spike-count relationship (Fig. 3B). These plots suggested the existence of a sublinear dependence between variance and mean firing rate across areas: VPL, areas 3b and 1 (30). Then, we pooled the Fano Factor for neurons from each area to compute their distributions (Fig. 3D). Importantly, both the median of each distribution ($F_{VPL}$=0.692;

$F_{Area\ 3b}$=0.731; $F_{Area\ 1}$=0.899) and the median of the Fano factor drop during the first half of the stimulus period ($F_{VPL}$ = 0.17; $F_{Area\ 3b}$ = 0.36; $F_{Area\ 1}$ = 0.66) showed again a hierarchical ordering across the three areas ($F_{VPL}$<$F_{Area\ 3b}$<$F_{Area\ 1}$). Overall, these results demonstrate that the stimulus-driven inter-trial variability also reflects a hierarchical ordering, showing that area 3b neurons lay in between the distributions associated with VPL and area 1, respectively.

Importantly, in both variability metrics, time constant decay and Fano Factor, there is a slight difference between VPL and 3b responses. By contrast, there is a notable difference between VPL and area 1. With the aim to examine whether differences are present at the firing rate level, we performed a classification procedure. By using a nonlinear dimensionality reduction technique based on the topological similarities between the firing rate profile (Methods (40, 41)) between areas we found no groups between VPL and 3b (data not shown) but larger differences between VPL and area 1 (Fig S2A). The dimensionality reduced data fed a non-linear support vector machine to find the best frontier between the activity of both areas (dashed line in Fig. S2A). The overall performance of the classifier was 66% ± 25% by using a cross-validation by using 20% of data in the testing epoch. Complementary, the population mean firing rate for amplitudes 0, 6, and 24μm is shown in Fig. S2B. In this, is notable a difference in the first ms of the stimulus (related to adaptation to stimulus), notably the fall in the activity in VPL is squared while in the area 1 is much abrupt. This procedure was employed recently to exhibit hierarchy differences between area 3b and 1 in a bimodal task (42). Importantly, to the best of our knowledge, this is the first evidence of the similarities in the activity of VPL and 3b and, at the same time, showing a clear hierarchy between 3b/VPL and area 1.

Additionally, we examined the relation between the intrinsic timescale and the Fano factor across neurons during stimulus-present trials (Fig S3) for VPL, areas 3b and 1. Note that both variability metrics ($\tau$ and Fano) exhibit an inherent correlation during the basal period. The correlation traces show a hierarchical decay across areas during the first half of the stimulus period (1.5-2s, VPL, $|\Delta\rho|$=0.03; area 3b, $|\Delta\rho|$=0.15; area 1, $|\Delta\rho|$=0.15). Hence, correlation between variability and intrinsic times diminish during stimulation. Further, area 1 displays the stronger decorrelation during stimulation, providing evidence that both sources of variability become uncoupled as the sensory information is projected from the thalamus to S1.

**Parallel feedforward pathways between VPL-area 3b and VPL-area 1 emerge during stimulus arrival.**

To estimate the directed information interactions across the studied areas during correctly detected trials (0-4s), we employed a methodology developed in previous works (20, 43). Briefly, we used a non-parametric method that measures the directed coupling between the simultaneously recorded spike trains of pairs of neurons in single trials and within slicing time windows of 0.25m. The method is illustrated in Fig. 4A. We first estimated delayed versions of the directed information-theoretic measure in both directions for every pair of neurons at the short time delays [0,2,4,…,20 ms]. To infer the significance of each estimation, we defined a maximizing-delay statistic and built the corresponding null distribution (Fig. 4A, middle). For each directional spike-train pair, the method assessed the significance of the statistic together with an unbiased estimation of the statistic value and the maximizing-delay (Fig. 4A, right). Spike-train pairs associated with significant estimators ($\alpha=0.05$) are referred to as Directional Information (DI) trials and will be represented for different experimental conditions as a percentage over the corresponding pairs and trials.

When studying DI for pairs of spike trains belonging to different areas along the somatosensory network VPL-3b-1, we assume that feedforward interactions are composed of VPL→area 3b, VPL→area 1, area 3b→area 1. Analogously, we assume that feedback interactions are composed of area 1→area 3b, area 1→VPL, area 3b→VPL. Moreover, we also considered bidirectional interactions (20). These types of interactions are mounted by bidirectionally coupled spike trains, that is, spike trains that are statistically coupled in both possible directions and usually occur at delay of 0ms (20). For the sake of clarity, Fig. 4B shows a schematic representation for the three types of interactions.

The estimation of interactions across inter-area pairs in VPL-3b-1 (Fig 4C) revealed that feedforward DI (blue curves) prevailed over feedback DI (red curves) during the stimulus period for both VPL-area 3b and VPL-area 1. In general, DI percentage after the stimulus period did not differ from pre-stimulus values, suggesting a stimulus-driven DI modulation in these areas. In both cases, the percentage of DI after stimulus onset (1.5-1.75s) significantly increased with respect to pre-stimulus values (VPL→area 3b, $P < 0.01$; VPL→area 1, $P < 0.01$). Thus, feedforward DI is comparable between VPL-area 3b and VPL-area 1. So, even if a hierarchical ordering could be established across these areas, the thalamic influence appears to be akin. For pairs area 3b-area 1, feedforward DI was also significant during the stimulus, however, were lower in magnitude in comparison to VPL feedforward interactions (Fig. 4C, right). In sum, these results suggest a parallel processing occurring at area 1 and area 3b being partially orchestrated by Thalamus (VPL).

In addition, the stimulus-driven nature of the described results was supported by the replication of Fig. 4C during correct rejected trials (without stimulation, Fig. S4). Indeed, Fig. S4A shows that there were no significant DI variations with respect to the baseline period during neither the possible stimulation window (44) nor any posterior task interval.

To further corroborate our results, we reproduced Fig. 4 using cross-correlation (CC) but following the same statistical analysis (surrogate testing, see Methods). Employing the lag, we were able to differentiate feedback, feedforward, and bidirectional CCs (Fig. S5A). Notably, Fig. S5C and E demonstrate that the interaction trends are preserved using CC. Nevertheless, CC falls short to capture the same percentages of interactions across areas as the DI during the stimulus period, which might reduce the statistical power in relating connectivity outcomes with variability measures.

Given that feedforward DI from VPL is analogous between area 3b and 1, it suggests that area 3b is not an indispensable relay in broadcasting the sensory information that arrived from VPL towards area 1. To further test this conjecture, we analyzed the feedforward DI delays associated to each inter-area (Fig S6). Importantly, the histograms displayed in Fig S6C showed that the feedforward median delay between VPL and area 1 ($\widehat{D} = 8$ ms) was incompatible with the addition of feedforward VPL→area 3b ($\widehat{D} = 6$ ms) and area 3b→area 1 ($\widehat{D} = 12$ ms). This result does not support the relay role hypothesis for area 3b. Further, it agrees with Fig. 4C (right) that shows that feedforward interaction between areas 3b and 1 was much less modulated than those originates from VPL. Moreover, only bidirectional area 3b↔area 1 DI appeared with akin modulation. These results suggest that simultaneous inputs might delivered to both areas that synchronize their activity. This means that information appears to be streamed in parallel pathways from VPL to areas 3b and 1.

Finally, we analyzed the dynamics of bidirectional interactions over the time course of the task (Fig 4C, orange). This coupling was shown previously to be a relevant feature for effective vibrotactile stimulus detection (20). Note that all areas interactions demonstrated a significant bidirectional DI modulation during the stimulus period. Our results demonstrate that a common synchronization among the three areas emerged during stimulation. This coordination across areas may play a mechanistic role in the transmission and stability of sensory information from VPL to S1 (20).

**Stimulus-driven intra-area interactions in area 1.**

Afterwards, we focus on the interaction within each area (VPL, 3b and 1). In this case, DI estimations could be of two types: 1) unidirectional interactions, in which spike-trains from the same area were coupled in one out of the two possible directions; 2) bidirectional interactions synchronization, when spike trains were coupled in both directions. Fig. 4D shows a schematic representation for both types of interactions. Fig. 4E shows the percentage of DI within each area across supra-threshold trials (Fig 4E). We found that the arrival of stimulus elicited only a significant unidirectional and bidirectional intra-area DI in area 1 ($P < 0.01$). Then, the intra-area interactions in VPL and area 3b were not significantly modulated during stimulation. We replicated these results for stimulus-absent trials (Fig S4B) to manifest the stimulus-driven effect observed in Fig. 4E.

In conclusion, the above results suggest different roles for the inter-area and intra-area interactions across these areas. Our data shows that upon stimulus arrival, VPL establishes comparable stimulus-driven feedforward DI to both somatosensory areas, leading to a rapid synchronization across the three areas (VPL↔area, 3b↔area 1) to strengthen these interactions. Further, area 1 concurrently exhibits a significant increase of local interactions (Fig. 4E right) that do not arise in either VPL (Fig. 4E left) nor area 3b (Fig. 4E middle). This increase (Fig. 4E right) is also concurrent to a deep decay of area's 1 Fano factor (Fig. 3A right). Therefore, area 1 is equipped with longer timescale, modulated stimulus-driven variability and displays stronger intra-area stimulus-driven interactions, essential features to integrate, transform and maintain sensory information.

**Relationship between Fano factor and thalamocortical interactions.**

In this section, we quantify the interplay between local variability measures and thalamocortical interactions. First, we computed the association between the Fano factor (Fig 3A) and the inter-area feedforward DI during the stimulus period. To illustrate the associations between Fano factor and feedforward DI, in Fig 5A-B we divided neurons from each area into two groups (low [light] and high [dark] Fano factor neurons) according to the area's median Fano factor during stimulation. Fig. 5A shows that, for VPL and area 1 especially, this division of neurons separated low and high Fano factor neurons throughout the entire task, which demonstrated that variability is a representative feature of each neuron. Afterwards, in Fig 5B we computed the percentage of feedforward DI from VPL (Fig. 5B left) to area 3b (Fig. 5B middle) and to area 1 (Fig. 5B right), for low and high Fano factor neurons. Indeed, we further illustrate that both the low variability group in VPL (Fig. 5B) and the high variability group in area 3b (Fig. 5C), yielded larger feedforward DI during the stimulus period ($P < 0.01$). In contrast, no directional information differences across groups were observed in area 1 (Fig. 5D and 5G). Then, the incoming feedforward DI to area 1 was totally uncoupled from the inter-trial variability throughout the entire task. Taken together, these findings report that variability may play a relevant role for establishing feedforward DI in VPL and area 3b, but this effect is diluted in area 1. In particular, the roles of VPL and area 3b variability were found to be the opposite when establishing feedforward connections with area 1. While feedforward DI to area 1 was sustained by low Fano factor neurons in VPL, that is, neurons with more capacity to encode stimulus information, it was driven in area 3b by neurons with the highest inter-trial variability (and hence, less neural coding capacity), thus providing further evidence that the area 3b →area 1 pathway might be less relevant for information processing.

To verify that the differences in DI observed across the two Fano groups could not be explained by firing rate changes, we also split the neural responses based on their mean firing rate (Fig. 5E-F). We found that this division of neurons did not generate significant differences throughout the stimulus period. The above results suggested a relation between the firing rate variability (but not the pure firing rate) of VPL and area 3b neurons and their feedforward associations with area 1. More precisely, it meant that the less variable the neurons of VPL were across trials, the more prone they were to establish significant feedforward interactions. Therefore, more reliable neurons in VPL (low Fano factor) tend to create significant feedforward information interaction. On the contrary, the firing rate variability of neurons in area 1 was unaffected by the incoming information from VPL, suggesting that the influence of feedforward inputs into the internal dynamics of area 1 was modest.

**Relationship between intrinsic timescales and inter and intra-area interactions.**

Subsequently, we study the relationship between thalamocortical DI and each area's intrinsic timescale (Fig 2). Unlike the Fano factor, the timescale was estimated once, during the pre-stimulus period. In this section, we included in the analysis the interaction between VPL and the whole S1 (VPL→S1). Analogous to Fig. 5, we estimated the association across neurons' timescales and their feedforward DI. We divided neurons from each area into two groups (low [light] and high [dark] $\tau$ neurons) according to each neuron's timescale with respect to the area's median value. For each group, we computed the percentages of intra-area DI for VPL (Fig. 6A, left), area 3b (Fig. 6B, left), and area 1 (Fig. 6C, top left). Concordantly, looking at the individual neurons, we observed a significant correlation between timescales and intra-area DI in area 1 (Fig. 6C, bottom left, *P<0.0001*). On the other hand, the influence of timescale on intra-area DI was not significant in VPL (Fig. 6A, left) and much more modest in area 3b (Fig. 6B, left). These findings support modelling works that establish a differentiated relationship between intrinsic timescales and intra-area recurrent interactions along the somatosensory pathway VPL →area 3b →area 1 (1).

Finally, we extended this analysis to the inter-area results. We split the neurons from each area (Fig. 6A VPL; Fig. 6B area 3b; Fig. 6C area 1) according to their timescale and for each group we calculated the feedforward DI from VPL→area 1 (Fig. 6A, right); from area 3b to area 1 (Fig. 6B, right); from VPL to area 1 (Fig. 6C, top right) and from area 3b to area 1 (Fig. 6C, bottom right). We did not find significant differences between feedforward DI from each group. Interestingly, in contrast to intra-area DI, the timescales from neurons in area 1 did not play a significant role regarding the incoming feedforward DI (Fig. 6C, right).

**DISCUSSION**

In brief, we show compelling evidence that a hierarchical order in the somatosensory network could be established by employing variability measures, intrinsic timescales and directed information (DI) measures across the different areas of the thalamocortical network. Although these measures quantify various aspects of the network dynamics, a hierarchical picture (1) could be established through assessing the associations between them. Therefore, this work establishes a relationship between these metrics, contributing thus to a better understanding of the role of neural variability in this network.

While time constants may play a relevant functional role to determine the period over which areas integrate their inputs (1, 21, 32), inter-area DI represents the degree of recurrent interplay. Here, we provide evidence that relates the timescales with recurrent interactions. As it is known, brain circuits require a diversification of recurrent connectivity and timescales during perceptual processes (45). Under this requirement, long time constants and strong recurrent interaction may promote the sensory transformation that depends on signal integration or its maintenance. However, long time constants, strong variability, and inter-area information flow, could be an impediment when generating a precise representation of the stimulus. Then, networks with more modest recurrent interaction are still necessary. This might yield a connection between functional and structural heterogeneities.

**Functional insights on the thalamocortical circuit**

Based on previous anatomical evidence, VPL is viewed as a first-order relay nucleus (11), where somatosensory information is conveyed to areas 3b and 1. Our results provide further evidence to support this idea by showing that the intra-area information flow within VPL is not modulated. Then, in contrast to areas 3b and 1, recurrent processing (intra-area interactions) is not modulated during the sensory input within VPL. In agreement with this result, autocorrelation decay is much smaller in VPL than S1, indicating a modest reverberation. Moreover, VPL shows smaller firing rate variability than both subareas of S1. Further, VPL neurons with smaller Fano Factor, tend to communicate more information to cortical neurons, which means that the VPL tendency to display a small variability, is related with its main goal, to transmit reliable information to the cortical somatosensory network.

Historically, VPL and S1 were considered as sensory areas, implicated in pure phase locking responses (3, 46, 47). Additionally, areas 3b and 1 were both considered as part of S1, with analogous responses and few or nonclear differences between them (48, 49). Further, recent anatomical evidence has suggested that both areas show akin amounts of synaptic projected from VPL (50). Here, we confirmed that during the perceptual detection of the vibrotactile stimulus, the amount of incoming feedforward interactions during the stimulus period is comparable in area 3b and area 1. Additionally, contrary to the hypothesis that propose area 3b as an information relay to area 1, feedforward interactions during stimulation were poorly modulated and mainly sustained by neurons in area 3b with large firing rate variability (hence, with low encoding capacity). Then, even if discrepancies were found between these areas, they are not established by differences in VPL inputs. One hypothesis is that they emerge because of structural differences in these networks (1). In relation to this conjecture, we identified a functional difference between these areas. Whereas neurons from area 3b exhibit similar variability properties to VPL such as small Fano factor decays during stimulus presentation and intrinsic timescales, area 1 yields much higher variability and time constant. This result might imply that one area is more equipped to represent VPL inputs, while the other to process and to integrate somatosensory information that has already reached the cortex. Importantly, even when area 1 exhibited a higher functional hierarchy than area 3b, advanced cognitive process such as decision making or crossmodal interactions emerge in later stages of the pipeline. In this direction, a recent study in a bimodal detection task suggest that both subareas are unimodal (42). Additionally, previous work has suggested that neurons from both subareas did not encode decision or choice (48). Then, even when there is a neural code transformation between 3b and area 1, these results indicate that higher abstract representations emerge at forward stages of sensory information processing. Remarkably, our results show that transformation between VPL to S1 and similarly, that from 3b to area 1 reported by previous studies is a continuous one (42) in consequence, there is no clusters of activity.

**Connecting neural variability and inter-and intra-area connectivity**

Although recording in isolated neurons exhibit stochastic variability, they appear much more stable than recording from brain areas (51). Then, a portion of the variability occurs from variations in the synaptic currents associated to each network (29, 30). Previous studies have suggested that such degree of variability at a single-neuron level increases with the stages of sensory processing, being lowest in the periphery and highest in cortical structures (30, 34, 38). Importantly, VPL, areas 3b and 1 displayed hierarchical difference between their variability during resting state and stimulation period. Notably, area 1 shows the highest variability during the basal period and the strongest decrease during the stimulation period. Additionally, during this period, the area 1 single-cell timescales decorrelate with their inter-trial variability and increase the relation with the intra-area interactions. Moreover, the inter-area feedforward input from VPL does not correlate with the area 1 variability. These findings exhibit dynamics differences between these areas. While variability in VPL and area 3b are affected by feedforward interactions, inputs to area 1 are diluted by network processing.

**Estimating statistical interactions via percentage of directional information**

It is important to emphasize that our analytical approach employs nonparametric significance tests to compute the percentage of significant DI to estimate the interaction between and across areas (52). Previous studies have shown that DI has a weak dependence on the firing rate of neurons. Indeed, it estimates the interaction across neurons in a quasi-orthogonal dimension to the neural activity. Therefore, unlike other multiple-trial methods (53), DI quantifies, in a single trial and for any given time window, how much information can be obtained from the recent delayed past of a given neuron about the present spike train of another, simultaneously recorded neuron. Here, we additionally provided a comparison of the DI method with surrogate-corrected cross-correlation (54, 55) applied to the same dataset, which highlighted the larger percentage of interactions that the DI method detected when the estimation signal-to-noise ratio was sufficiently high (stimulus period). Finally, we remark that our results rely on a correlation analysis between functional interactions estimated a single-trial level (and averaged across trials) and variability measures estimated at a multiple-trial level, thus adding additional robustness to potential sample biases.

**Limitations of the study**

Some of the main limitations of our study are related to the experimentally challenging protocol to have one neuron from a thalamic nucleus to be simultaneously recorded with other neurons from cortical areas. This lowered the number of neurons that could be simultaneously recorded from the three areas (VPL, and area 3b and 1) and prompted us to control their inherent neuronal variability via the definition of appropriate selection criteria prior to perform each initial independent analysis (timescale, Fano factor and DI). To analyze inter-area comparisons that included VPL we had to focus on analyzing interactions across pairs of neurons. Because of this circumstance, we chose a pairwise measure that does not take into account the activity of more than two neurons simultaneously (56, 57). Second, our experimental protocol relies on recordings of neurons in VPL, and area 1 and 3b of S1, and hence, the effects of subcortical areas as ventroposterior inferior thalamus, basal ganglia, or secondary somatosensory cortex are out of the scope of this study. In addition, our recording protocol does not consider the cortical layers of the recorded neurons, which is needed for a deep characterization and understanding of the cortico-cortical and cortico-thalamic interactions. Furthermore, activity inactivation techniques would be desirable to leverage our results and hypothesis, however changes in the network due to the inactivation of the whole part would result in changes in no obvious ways at the interaction levels and would even obscure some interpretations. Additionally, in this article we do not consider other interaction levels as could it be: interactions between fields and spikes or even field-field interactions (58). These methodological procedures will be addressed in our future research. Further, we analyzed the comparison of the results of our measure with cross-correlation. However, it remains to be examined how the applied methodology compares with other linear methods relying on kernel couplings (59, 60). Finally, we did not propose here a model explaining the reported differential integration of VPL and cortical inputs by area 1. Yet, we believe that deep convolutional neural networks that hierarchically transform information across areas (61, 62) could serve for this purpose in future follow-up works.

One important question for future research is how oscillation signals like the local field potentials (LFP) are affected by variability, timescales, and local and inter-area interactions. Is it possible to observe these differences in hierarchy looking at the inter and intra-area oscillatory coherence? Furthermore, future experiments involving multi electrodes recording are essential to clarify the role of the different cortical layers (63). Are there any differences in variability and time constant across each cortical column? How do the intra, feedforward and feedback information flow change across the different cortical layers? Is it possible to establish a hierarchy of processing across these layers?

**Concluding remarks**

To conclude, we would like to point out that the synergy between feedforward and feedback interactions and inherent dynamical features like the ones computed here, shape the function of each brain network. Diversification of structural and dynamics features among cortical and subcortical networks may be central to discerning brain function and computations during cognitive processes.

**ACKNOWLEDGMENTS**

We thank Hector Diaz for his technical assistance. A.T.C was supported by the Spanish national research project (ref. PID2020-119072RA-I00 /AEI/10.13039/501100011033) funded by the Spanish Ministry of Science, Innovation, and Universities (MCIU) and by the Bial Foundation grant 106/18. G.D. was supported by the Spanish national research project (ref. PID2019-105772GB-I00 /AEI/10.13039/501100011033) funded by the Spanish Ministry of Science, Innovation, and Universities (MCIU). This work was supported by grants PAPIITIN205022 from the Dirección de Asuntos del Personal

## METHODS

### Detection task

The detection task and neural recordings follow along the same lines as (20, 25, 64). Vibrotactile stimuli were delivered to the skin of the distal segment of one digit of the restrained hand, via a computer-controlled stimulator (BME Systems, MD; 2-mm round tip). The initial probe indentation was 500 μm. Vibrotactile stimuli consisted of trains of 20 Hz mechanical sinusoids (20 ms duration), with amplitudes of 1-34 μm (Fig. 1A). These were interleaved with an equal number of trials where no mechanical vibrations were delivered to the skin (amplitude = 0). A trial began when the probe tip (PD) indented the skin of one fingertip of the restrained, right hand, upon which the monkey placed its free, left hand on an immovable key (KD). After a variable pre-stimulus period (1.5-3 s), a vibrotactile stimulus could be presented or not (0.5 s). Note that the possible period of stimulation lasts 1.5s longer than the stimulus, as can be seen in the differences in shadow areas in Fig. 1E (top) and Fig. 1E (bottom). After a fixed delay period (3 s), the stimulator probe was lifted off from the skin (PU), indicating to the monkey that it could initiate the response movement (KU) to one of two buttons (PB). The button pressed indicated whether the monkey felt the stimulus (henceforth referred as 'yes' and 'no' responses, respectively). They were rewarded with a drop of liquid for correct responses. Psychometric detection curves were obtained by plotting the proportion of 'yes' responses as a function of the stimulus amplitude (left panel of Fig. 1B). Depending on whether the stimulus was present or absent and on the behavioral response, the trial outcome was classified as hit, miss, false alarm or correct rejection (right pane of Fig. 1B). Monkeys were handled according to the institutional standards of the National Institutes of Health and Society for Neuroscience. All protocols were approved by the Institutional Animal Care and Use Committee of the Instituto de Fisiología Celular of the National Autonomous University of Mexico (UNAM).

### Recordings

Neuronal recordings were obtained with an array of fourteen independent, movable microelectrodes [2–3 MΩ; (4, 25, 64)] inserted into S1 (seven), medial to the hand representation in such a way that allowed us to lower the microelectrodes into the VPL (seven). This maximized the probability of mapping the hand area in the VPL. At the end of each penetration, microlesions were made by passing 5-10 µA through the tip of the microelectrodes for 5s, to aid reconstruction of the penetration. Neurons were classified as belonging to areas 3b and 1 according to previous studies of the cytoarchitecture of the monkey postcentral gyrus. One array was inserted into S1 normally to the cortical surface (cyan spot on the figurine of the left panel of Fig. 1C), in the cutaneous representation of the fingers (areas 1 or 3b; middle panel of Fig. 1C). The other array was located lateral and posterior to the hand's representation (green spot on the left panel of Fig. 1C) of the Ventral Posterior Lateral thalamus (VPL). We located the microelectrodes in the cutaneous representation of the fingers in the VPL. Recordings were performed contralateral to the stimulated hand (right) and ipsilateral to the responding hand (left). Each recording began with a mapping session to find the cutaneous representation of the fingers in VPL. Subsequently, we mapped neurons in S1 sharing receptive fields with the neurons of VPL (Fig. 1D). Additionally, neurons from S1 were classified into areas 1 (n=336) and 3b (n=84). The neuronal signal of each microelectrode was sampled at 30 kHz and spikes were sorted online. A more extensive description of the task and recording procedure can be found in previous publications (4, 25, 64). Here, we report data from multiple recording sessions during which spikes were obtained. For the experimental condition, we recorded neurons from VPL (n=96) and S1 (n=420) during 85 sessions with 100−140 trials per session. The number of pairs recorded simultaneously organized per area is: 142 for area1-area1, 57 for area1-3b, 75 for 3b-3b, 71 for VPL-area1, 52 for VPL-3b and 119 for VPL-VPL. All activity is aligned to the minimum duration of the variable period, 1.5s before stimulus onset. During neural recording, the location and size of the receptive field were estimated by stimulating the glabrous skin of the stimulated hand. Then by keeping only neurons with small receptive fields (which could be shared with neurons in VPL).

**Intrinsic timescales estimation**

We computed the intrinsic timescales following the same method described in (32, 33) applied at single neuron level. Let us first model the time-binned spike counts of a spike train as a doubly stochastic process in which each time bin is of duration $\Delta$. Then, the autocorrelation *R(.)*, being a function of the time lag *k$\Delta$ (k=1,2,…)* between pairs of time bins indexed by their onset times $i\Delta$ and $j\Delta$ ($k=|i-j|$), can be approximated by the exponential decay function:

$$R(k\Delta) = A\left(e^{-\frac{k\Delta}{\tau}} + B\right),\qquad \text{[eq. 1]}$$

where $A$ is a multiplying constant, $\tau$ is the decay rate constant measuring the intrinsic timescale, and $B$ accounts for the contribution of timescales longer than the observed windows.

We computed the autocorrelation function $R(.)$ in our datasets along similar lines of previous works (32–34). First, from every single spike train, we obtained a spike-count time series binned in steps of $\Delta$ ms and windows widths of $W_b$ ms during a baseline period of 1.5s following the KU event and prior to stimulus onset. Then, in every neuron, for each pair of bins ($i\Delta$ and $j\Delta$, $i,j=1,2,…,i\neq j$) we computed the Pearson correlation between the spike count values at bins $i\Delta$ and $j\Delta$, respectively, over all recorded trials. Finally, to obtain the autocorrelation function $R(k\Delta)$, Pearson correlation coefficient values were accumulated and averaged over at each time bin difference $k\Delta$ ($k=|i-j|$). The maximum time difference is $D_{max}=k_{max}\,\Delta$. To obtain the above autocorrelation function for every single neuron, we accumulated and averaged the correlation coefficient over all time bin pairs at each time bin difference $k\Delta$. Then, the autocorrelation functions were fit by the exponential decay function [eq. 1] using nonlinear least-squares fitting via the Levenberg-Marquardt algorithm. For each brain area under study, we only selected neurons whose fitting outcome was sufficiently good base on either the fitted parameters converged to a solution, or the residual difference was less than a specified tolerance ($10^{-6}$) over a maximum number of iterations ($500$). In the following we call these neurons *well-fitted neurons*. As a result, we obtained an estimation of the intrinsic timescale $\tau$ per well-fitted neuron.

To obtain an autocorrelation function at the population level, we accumulated correlation coefficient estimates over all time bin pairs and well-fitted neurons within an area, averaging them at each time bin difference $k\Delta$. After undergoing the fitting procedure, we obtained an estimation of the intrinsic timescale $\tau$ per area. In parallel, to assess the robustness of this estimation, we computed the autocorrelation matrix for each well-fitted neuron separately and averaged the matrices across neurons. Then, we defined the autocorrelation function at each time bin difference $k\Delta$ by averaging the coefficient estimates over the corresponding time bin pairs in this matrix. This alternative method yielded approximately the same quantitative results, thus validating the initial estimation.

In agreement with a previous work (33), we applied the above procedures with steps of $\Delta=20ms$, windows widths of $W_b=40ms$, and maximum time difference of $D_{max}=500ms$. The results did not change qualitatively for variations ±10% of $\Delta$ and $W_b$. In our dataset, the percentage of well-fitted neurons per area was 32% in VPL (n=21), 57% in area 3b (n=31), and 63% in area 1 (n=87). Finally, confidence intervals for the intrinsic timescales were estimated using non-parametric bootstrapping (65).

**Classification of stimulus-responsive neurons for Fano factor and DI estimation**

During neural recording, the location and size of the receptive field were estimated by stimulating the glabrous skin of the stimulated hand. We focused our analysis on neurons with small receptive fields located in the stimulation zone. This allowed us to concentrate on neurons more likely to be the most responsive to stimuli, resulting in more reliable Fano factor and Directional information measurements. From this initial ensemble, a selection was made according to the following procedure. Each neuron's firing rate responses were subjected to linear regression analysis, comparing their activity upon initial stimulus presentation with the stimulus amplitude. Neuron's quality response was selected based on a threshold of 0.35 for the coefficient. Following the initial filtering process, we examined the individual neurons and neuron pairs. The presented tables illustrate the distribution of these neurons/neuron pairs across various areas and monkeys.

|       | VPL | area 3b | area 1 |
|-------|-----|---------|--------|
| M1    | 49  | 15      | 57     |
| M2    | 16  | 0       | 0      |
| M3    | 0   | 36      | 37     |
| M4    | 0   | 3       | 43     |
| Total | 65  | 54      | 137    |

**TABLE 1**. Individual neuron distribution after the quality response's classification.

|       | area 1 - area 1 | area 3b - area 3b | VPL - VPL | area 3b - area 1 | VPL - area 1 | VPL - area 3b |
|-------|-----------------|-------------------|-----------|------------------|--------------|---------------|
| M1    | 28              | 5                 | 5         | 6                | 64           | 13            |
| M2    | 0               | 0                 | 6         | 0                | 0            | 0             |
| M3    | 35              | 35                | 0         | 25               | 0            | 0             |
| M4    | 59              | 1                 | 0         | 4                | 0            | 0             |
| Total | 122             | 41                | 11        | 35               | 64           | 13            |

**TABLE 2**. Pairs of simultaneously recorded neuron distribution after the quality response's classification.

**Fano factor estimation**

Let $N(w)$ be a counting process describing the number of spikes in a time interval $(0,w)$, of length $w>0$, in which the time zero is conventionally set a priori. The Fano factor is a measure of the variability of $N(w)$ defined as the variance to mean ratio of the number of spikes in a time window of a length $w$ (30, 31, 37):

$$F(w) = \frac{Var(N(w))}{E(N(w))}, w > 0 \qquad [\text{eq. 2}]$$

The Poisson process is an example of a counting process for which the above variance-to-mean ratio equals one. Since Poisson processes have traditionally been a popular model to model stimulus-evoked spike counts (29), the Fano factor has been employed in many studies to characterize the degree of overdispersion of the observed neural firing rate with respect to the Poisson distribution. Hence, in practice, a Fano factor significantly larger or smaller than 1 is an indicator that the variability of the observed counting process is larger or smaller than the variability of a Poisson process, respectively.

In practice, the Fano factor for a single neuron in a limited time window is estimated based on $n$ independent observed spike counts during repeated trials of a fixed experimental condition (e.g., trials in which a vibrotactile stimulus of the same amplitude is delivered). The standard estimator of the Fano factor is then based on the ratio of unbiased estimators of the variance and mean of the *n* spike counts (see for instance (31), eq. 20). In our study, for every neuron and task interval of 250 ms, we estimated the Fano factor over *n* hit trials at a fixed amplitude and averaged the Fano factor estimates across amplitude values. For each Fano factor estimation, we selected *n* to be equal or larger than 5 trials. We removed Fano factor estimates lying *3* standard deviations above and below the median value prior to perform the stimulus-amplitude average. On the other hand, during stimulus-absent trials, we estimated the Fano Factor from correctly rejected trials (*n≥5*). Hence, during both conditions (hit and correct rejected trials), we obtained one Fano factor estimate per neuron and task interval.

Although the Fano factor has been popularly used in many neuroscience studies (36, 38, 39, 66), recent works have showed that the estimation of the *F(w)* might be influenced by the neuronal firing rate and advocate for unbiased definitions of neural variability (30, 31, 37). However, in our study we obtained one variability measure per neuron, and the resulting values only showed a significant dependence on the mean firing rate in VPL, which was of negative sign and temporally localized outside the stimulus period. Consequently, we discarded introducing any firing-rate correction in the analysis.

**Directional Information (DI) estimation**

The method used to estimate directional interactions at a single-trial level was also described in detail in (20). We estimated directional information between every neuron pair within a population using a Bayesian estimator of the directed information-theoretic measure (43) between a pair of discrete time series that were assumed to be generated according to a Markovian process. In more specific terms, for a pair time series $(x^T, y^T)$ of length T, where $x^T = (x_1, \ldots, x_T)$ and $y^T = (y_1, \ldots, y_T)$, a time delay $D \geq 0$, and Markovian orders equal to $M_1 > 0$ and $M_2 > 0$, respectively, the directed information-theoretic measure between the underlying stationary processes of $x^T$ and $y^T$, i.e., $(X, Y)$, is estimated through the formula:

$$\hat{I}_D(X \rightarrow Y) \triangleq \frac{1}{T}\sum_{t=1}^{T}\sum_{y_t} \hat{P}\left(Y_t = y_t \middle| X_{t-D-M_2}^{t-D} = x_{t-D-M_2}^{t-D}, Y_{t-M_1}^{t-1} = y_{t-M_1}^{t-1}\right) *$$

$$log \frac{\hat{P}\left(Y_t = y_t \middle| X_{t-D-M_2}^{t-D} = x_{t-D-M_2}^{t-D}, Y_{t-M_1}^{t-1} = y_{t-M_1}^{t-1}\right)}{\hat{P}\left(Y_t = y_t \middle| Y_{t-M_1}^{t-1} = y_{t-M_1}^{t-1}\right)},$$

[eq. 3]

Equation 3 quantifies the information that the past of $X^T$ at delay D, i.e., $X_{t-D-M_2}^{t-D}$, has about the present of $Y^T$, i.e., $Y_t$, given the most recent part of $Y^T$, i.e., $Y_{t-M_1}^{t-1}$. This estimator is consistent as long as the two neuronal time series $(X^T, Y^T)$ form a jointly stationary irreducible aperiodic binary Markov process with a certain maximum order. Prior to estimating the directed information-theoretic measure, we preprocessed our data as follows. For a single trial, we first binarized spike-train trials using bins of 1ms (mapping 1 to each bin with at least one spike and 0, otherwise). Second, in stimulus-present trials, we removed the variable-time pre-stimulus period in every trial and aligned all trials to the stimulus onset time. In contrast, in stimulus-absent trials, we aligned the trials to the probe down event (PD). We then divided each trial time series into sixteen non-overlapping task intervals of 0.25s (250 bins). At each task interval, the spike train was assumed to be generated by a random process that satisfied the estimator requirements with a maximum memory of 2ms ($M_1 = M_2 = 2$ bins) both for the joint and the marginal spike-train processes. Under the estimator requirements, it can be easily checked that the directed information-theoretic measure is asymptotically equivalent to the transfer entropy measure in the limit of the time-series length. Finally, among those trials, we ran the delayed directed information-theoretic measure estimator at time delays D=0, 2, 4, 6, 8, 10, 12, 14, 16, 18, 20 ms.

We dealt with the multiple test problem over delays by using the maximum directed information-theoretic measure over all preselected delays as a test statistic:

$$I_{STAT}(X \rightarrow Y) \triangleq \hat{I}_D(X \rightarrow Y) \quad \text{[eq. 4]}$$

To assess the significance of the above statistic (Eq. 4), we used a Monte-Carlo permutation test. In this test, the original (i.e., non-permuted) estimation was compared with the tail of a distribution obtained by performing 20 equally spaced (to maximize independent sampling) circular shifts of the target spike train $Y^T$ within the range [50, 200]ms and computed the corresponding P-value. Hence, the significance test provides three outputs: the significance assessment (0/1), the statistic value and the maximizing delay $\widehat{D}$. In particular, any spike-train pair during a trial is considered to convey directional information (DI) at a given task interval if the corresponding test yields significance. The software to perform such estimations is available as a package ("DI-Inference") in both Matlab (https://github.com/AdTau/DI-Inference) and Python (https://github.com/mvilavidal/DI-Inference-for-Python) languages. The package includes an exemplary script with simple simulated toy models to evaluate the method's performance (67).

**Inter-area and intra-area interaction types**

As in (20), when studying DI interactions for two simultaneous spike trains at a given time interval (e.g., 0.25 s) one may consider three disjoint cases: the spike trains are coupled in only one direction, in only the opposite direction or simultaneously coupled in both directions. In principle, these three cases correspond to neurons in each pair taking three different roles: driver, target, or both. According to this notion, we classified DI estimates per trial by pairing the role (driver, target) of each neuron with its location according to the presumed directionality of the somatosensory pathway VPL →area 3b→area 1. In short, we defined as feedforward interactions the DI estimates obtained from genuine driver-target pairs (VPL, area 3b; VPL→ area 3b), (VPL, area 1; VPL→ area 1), and (area 3b, area 1; area 3b → area 1). Similarly, we defined as feedback interactions the ones obtained from the genuine driver-target pairs (area 3b, VPL; area 3b → VPL), (area 1, VPL; area 1→VPL), and (area 1, area 3b; area 1 → area 3b). Finally, pairs where both neurons were simultaneously drivers and targets were labelled as bidirectional interactions (VPL↔ area 3b, VPL ↔ area 1, area 3b ↔ area 1). In contrast, for intra-area interactions, since potential target or drivers were a priori functionally similar, we grouped all non-bidirectional cases as unidirectional interactions (e.g., area 1→area 1) and separated them from the bidirectional type. (e.g., area 1↔ area 1).

**Comparison with linear connectivity (cross-correlation)**

The reproduction of the DI results (Fig. 4) with the cross-correlation measure (Fig. S4) was conceptually the same as previous works based on correlation measures with jitter correction (54, 55). However, in this case, for the sake of a fair comparison between both measures, the significance testing followed the same procedure used for the DI. Hence, for each possible directed (ordered) spike train pair, the cross-correlation was computed over the range [0,2,..,20]ms and its significance was assessed using the same circular-shift surrogates. Then, directionality was assigned using the same rules used for the DI (See the section above). Finally, the outcomes of this analysis were represented as the percentage of significant cross-correlation estimates of each type (e.g. feedforward, feedback, etc.) over all trials.

**Statistical analysis: Quantification of significant effects on DI percentages**

The main metric used in Fig. 4 was obtained by aggregating each DI type (feedforward, feedback, and zero-lag interactions) over all neuron pairs and trials at individual task intervals and computing their percentage over the total amount of trials. Specifically, the stimulus-driven change in the percentage of DI illustrated in Fig. 4, was tested as an unpaired comparison using a non-parametric test for correlated samples (68) relying on Cohen's effect size (Cohen's H; (69)) statistic, which measures the distance between two proportions $p_1$ and $p_2$ as:

$$H^{\text{unpaired}}(p_1, p_2) = 2\left(\arcsin \sqrt{p_1} - \arcsin \sqrt{p_2}\right)$$

[Eq. 5]

The use of this statistic allows to straightforwardly quantify the size of any significant effect by comparing its value with standardized thresholds (H = 0.2, small effect size; H = 0.5, medium effect size; H = 0.8 large effect size), thus avoiding sample size biases. Non-parametric tests for correlated samples were performed through 1000 group-based permutations (68) where groups were defined to be single trials and group sample sizes were maintained in each permutation. Thus, our analysis avoided introducing any statistical bias to the sampled reference distribution.

**Correlation of Fano factor and intrinsic timescales with percentage of DI**

To assess the potential association between spike-train directionality across pairs of somatosensory areas and local neuronal variability, we correlated the percentage of the DI with the intrinsic timescale and the average Fano factor across all neurons of each area using Spearman's rank-order correlation ($\rho$). In this analysis, we mainly focus on the percentage of feedforward information during the first half of the stimulus period (0-250ms). This choice was made because feedforward information was shown to be largely modulated by the stimulus presence in real data and the DI in general was shown to be sufficiently robust to additive noise at the firing rates of the stimulus period in simulated models (See Ref. (20), Fig. S5). To compute the correlation ($\rho$), we obtained one directionality value per neuron by averaging the percentage of DI over its recorded pairwise interactions. By doing so, we avoided introducing repeated measures per neuron in the analysis.

For both, the intrinsic timescales and the Fano factor computed during the stimulus interval, we correlated the corresponding variability metric ($\tau$ or Fano) obtained for every single neuron with the percentage of incoming/outgoing feedforward information. We particularized the above analysis within each area (VPL, area 3b and area 1) leading to three distinct computations: the correlation of $\tau$/Fano with the percentage of (outgoing) VPL-S1 feedforward information across VPL neurons; the correlation of $\tau$/Fano with the percentage of (incoming) VPL-Area 3b feedforward information across area 3b neurons, and the correlation of $\tau$/Fano with the percentage of (incoming) VPL-area 1 feedforward information across area 1 neurons.

**Neural activity classifier**

To assess the differences between area's activity we developed a classifier by using a non-linear support vector machine (nSVM). This classifier received as input the 2d-reduced firing rate. This reduction was performed by a non-linear dimensionality reduction algorithm called UMAP (Uniform Manifold Approximation and Projection) (70). UMAP weights the local and global similarities by constructing a graph and after collapsing it. To build such a graph we selected a Euclidean distance. After the dimensionality reduction epoch, we trained and tested the nSVM by using a cross validation scheme with 25% of the data as testing and 75% as training. These codes were based on the Python Scikit-learn suite.

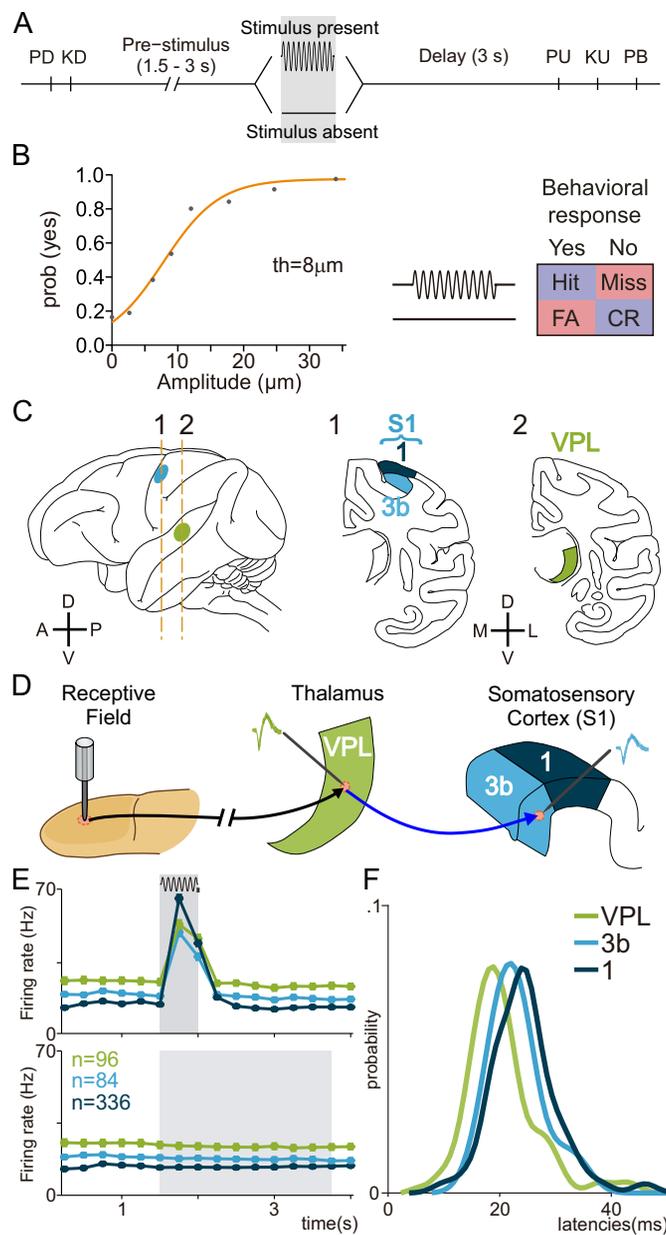

**Figure 1: Detection task, psychophysical performance, recording sites, and neuronal responses during the task.** (A) Vibrotactile detection task. Trials began when the stimulator probe indented the skin of one fingertip of the monkey's restrained right hand (probe down, PD); the monkey reacted by placing its left, free hand on an immovable key (key down, KD). After a variable prestimulus period (1.5 to 3 s), a vibratory stimulus of variable amplitude (1 to 34 μm, 20 Hz, 0.5s duration) was presented on one half of the trials; no stimulus was presented on the other half of the trials. Following the stimulus presentation period (whether the stimulus was present or not), the monkey waited for 3 s until the probe was lifted off from the skin (PU, probe up), then the animal releases the key (KU, key up) and pressed one of two push buttons (PBs) to report whether the stimulus was present (lateral button) or absent (medial button). Stimulus-present and stimulus-absent trials were randomly interleaved within a run. (B, Left) Mean psychometric function depicting the probability of the monkey's reporting yes (presence) as a function of the stimulus amplitude (th = 8 μm, detection threshold). (B, Right) Behavioral responses depending on the stimulus presence (Hit or Miss) or stimulus absence (CR, correct rejection; FA, false alarm). (C) Recording sites in the thalamus ventral posterior lateral (VPL) nucleus (green) and in areas 1 (dark blue) and 3b (light blue) of the primary somatosensory cortex (S1). (D) Scheme depicting how the neural activity from single neurons in the VPL and S1 (3b or area 1) sharing the same cutaneous receptive field was simultaneously recorded during the detection task. (E) Mean firing rate for the simultaneously recorded VPL (n = 96), Area 3b (n = 84) and Area 1 (n=336) neurons supra-threshold hits (top) and correct rejections (CR, bottom). The timescale is aligned to the minimum variable period, 1.5s before stimulus onset. Grey rectangle (top) represents stimulation period, while grey rectangle (bottom) represents the possible period of stimulation, according to task design. (F) Probability density of the neuronal response latencies with respect to stimulus onset during hit trials.

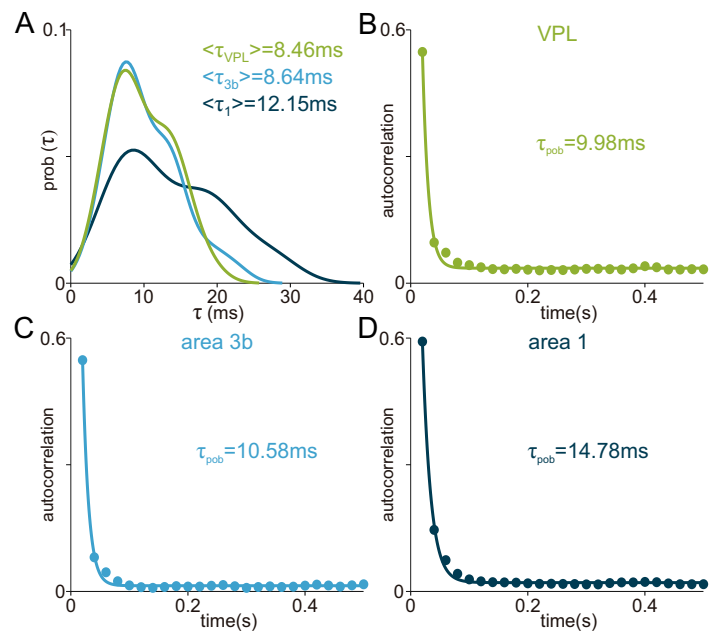

**Figure 2: Areas' intrinsic timescales are hierarchically ordered across the somatosensory network.** Autocorrelation functions were computed on the spiking activity recorded during the basal period (1.5s) before stimulus onset. An exponential decay function was fit to the resulting autocorrelation function. Confidence intervals for the decay rate parameter intrinsic timescale ($\tau$) were estimated through non-parametric bootstrap. (A) Distribution of neuronal $\tau$ in each area (green, VPL, n=21, median $\tau$=8.46ms; light blue, area 3b, n=31, median $\tau$=8.64ms; dark blue, area 1, n=87, median $\tau$=12.5ms.). (B-D) Population autocorrelation and fitted exponential decay function by each area. Colored filled dots represent the average autocorrelation values for each time bin difference (across pairs and neurons). Solid lines represent the fitted exponential decay function. (B) VPL (green, $\tau$=9.98±2ms). (C) Area 3b (light blue, $\tau$=10.11±2ms). (D) Area 1 (dark blue, $\tau$=14.06±3ms).

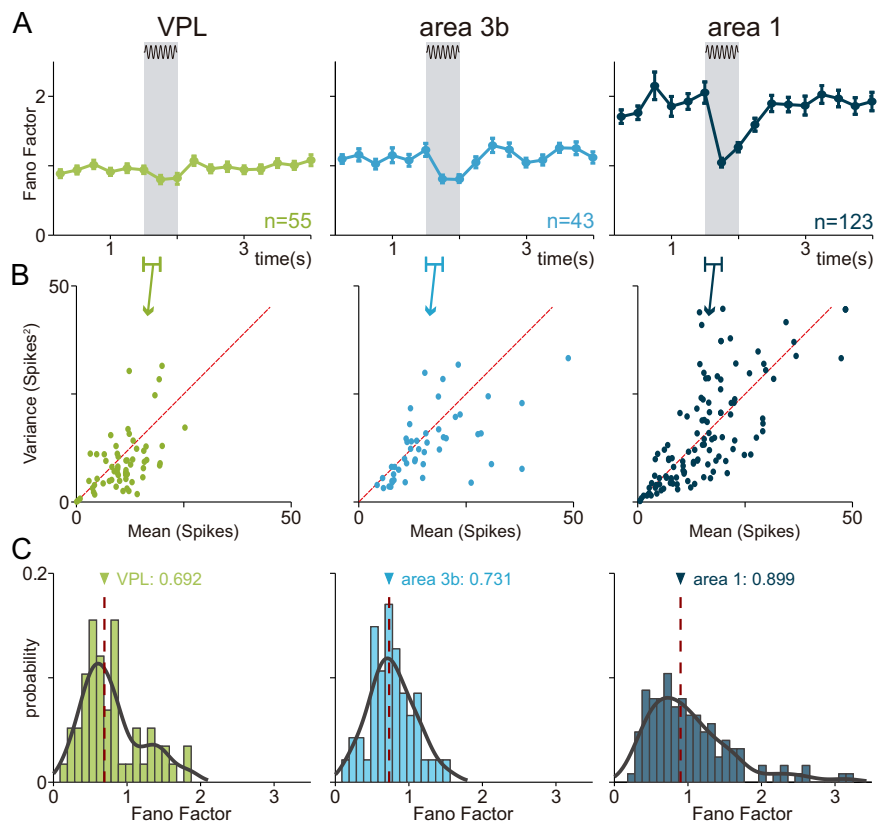

**Figure 3: Fano factor is hierarchically distributed across the somatosensory pathway.** (A) Time-varying average Fano factor during hit trials for each area (VPL, n=55 neurons; area 3b, n=43 neurons; area 1, n=123 neurons). Values for each neuron are computed and averaged over all amplitudes with enough hit trials (≤5). Error bars denote the SEM with respect to the mean over neurons in each area. Shaded area highlights the stimulus period. (B) Relationship between the average and the variance of the number of spikes during stimulus period intervals. Each dot corresponds to a different stimulus amplitude for each neuron The red dashed line plots the identity straight line (mean=variance) that should follow a Poisson distribution. (C) Histograms and estimated probability density of Fano factor per neurons during the stimulus period.

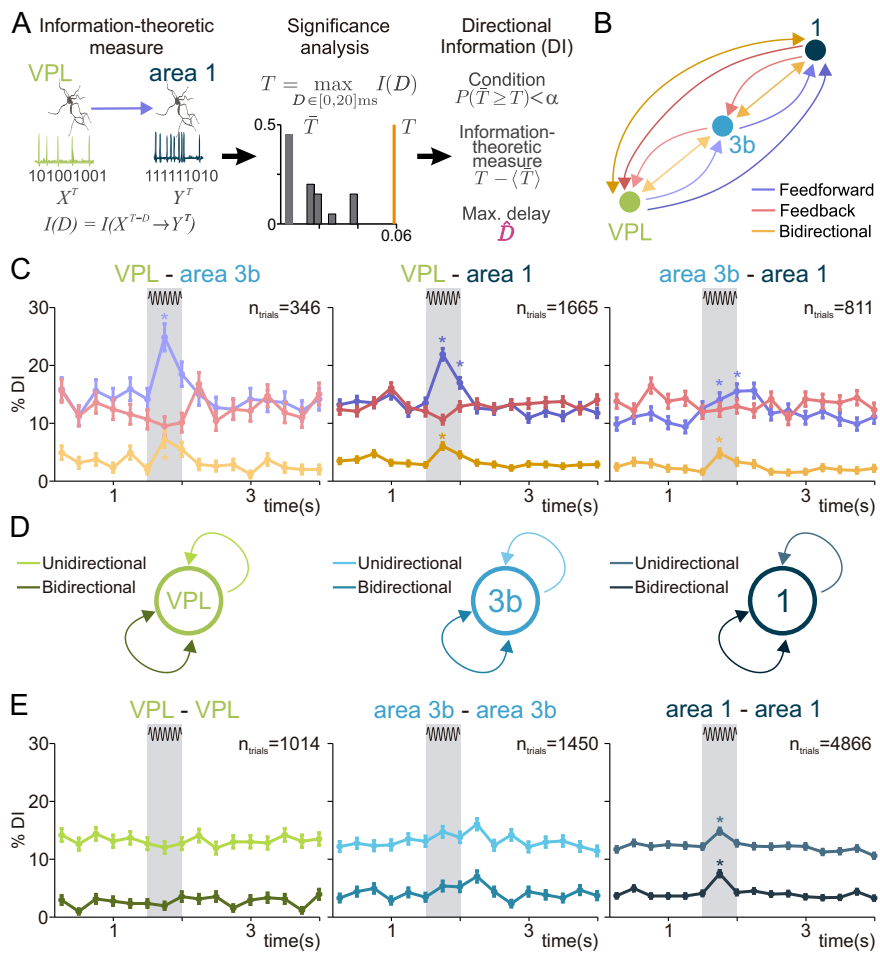

**Figure 4: Inter and intra-area directional interactions in the thalamocortical network.** (A) Sequential scheme representing the method to infer DI at single trial level. Left: information-theoretic measure is estimated between single-trial spike trains of the simultaneously recorded neurons in VPL and area 1 for delays (0,2,4, ⋯ ,20) ms. Middle: significance is locally determined via non-parametric testing ($\alpha=0.05$) of a maximizing-delay statistic. Right: every significant trial ($P<\alpha$) is denoted as Directional Information (DI) trial and is associated with an unbiased value and a delay. (B) Scheme showing feedforward, feedback and bidirectional interactions across VPL, area 3b and area 1. The arrows connecting VPL and area 1 are slightly darker colors to indicate the relationships in (C). Darker colors represent direct interactions without passing through area 3b as an intermediary. (C) Percentage of DI trials for feedforward (blue), feedback (red) and bidirectional (orange) interactions are shown along the task during supra-threshold hits. Left: VPL-area 3b neuron pairs (n=346 trials). Middle: VPL-area 1 pairs (n=1665 trials). Right: area 3b-area 1 pairs (n=811 trials). Error bars denote the SEM. Asterisks denote significant differences ($P < 0.01$) between a pre-stimulus baseline (1s) and stimulus periods with sufficiently high effect size ($H \geq 0.1$). H (Cohen's H) measures the effect size of each DI type: VPL→area 3b, $H=0.26$; VPL↔area 3b, $H=0.17$; VPL→area 1, $H=0.22$; VPL↔area 1, $H=0.12$; area 3b→area 1, $H=0.1$; area 3b↔area 1, $H=0.11$ (0-0.25s of stimulus period). Shaded area highlights the stimulus period (500 ms). (D) Scheme showing unidirectional and bidirectional interactions within VPL (left), area 3b (middle) and area 1 (right). (E) Percentage of DI trials for unidirectional (lighter color) and bidirectional interactions (darker color) are shown throughout the task during supra-threshold hit trials. Error bars denote the SEM. Asterisks denote significant differences ($P < 0.01$) between a pre-stimulus baseline (1s) and stimulus periods with sufficiently high effect size ($H \geq 0.07$). Left: VPL-VPL pairs (n=1014 trials). Middle: area 3b-area 3b pairs (n=1450 trials). Right: area 1-area 1 pairs (n=4866 trials). H (Cohen's H) for each significant DI type during first half of stimulation: area 1→area 1, $H=0.075$; area 1↔area 1, $H=0.15$.

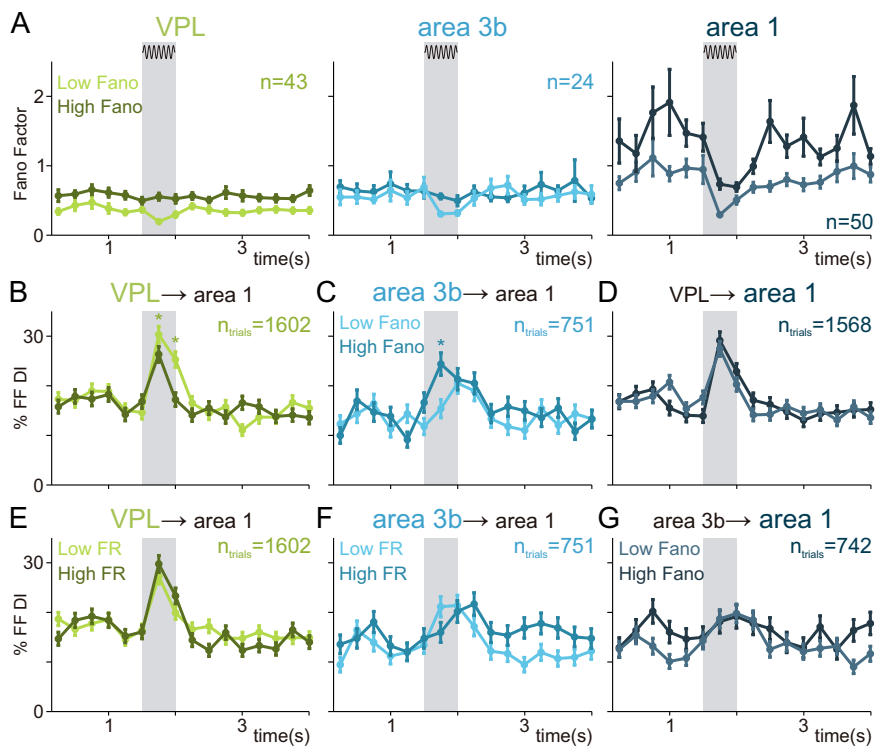

**Figure 5: Feedforward interactions are differentially related to the Fano Factor in each area.** (A) Average Fano factor during the task in VPL (left), area 3b (middle) and area 1 (right) as a function of two groups of neurons (low [light-colored] and high [dark] Fano factor neurons). These groups are split by the population median during the stimulus. (B)-(C) Percentage of feedforward interactions from VPL to area 1 (B) and from area 3b to area 1 (C) during the task (0-4s) for two groups split by VPL (B) and the area 3b (C) Fano factor median during the stimulus, respectively. (D) Percentage of feedforward interactions from VPL to area 1 during the task (0-4s) for two groups split by the area 1 Fano factor median during the stimulus. (E)-(F) Percentage of feedforward interactions from VPL to area 1 (E) and from area 3b to area 1 (F) during the task (0-4s) for two groups split by VPL (E) and the area 3b (F) firing rate median during the stimulus, respectively. (G) Percentage of feedforward interactions from area 3b to area 1 during the task (0-4s) for two groups split by the area 1 Fano factor median during the stimulus. Error bars denote the SEM. Shaded areas highlight the stimulus period (1.5-2 s).

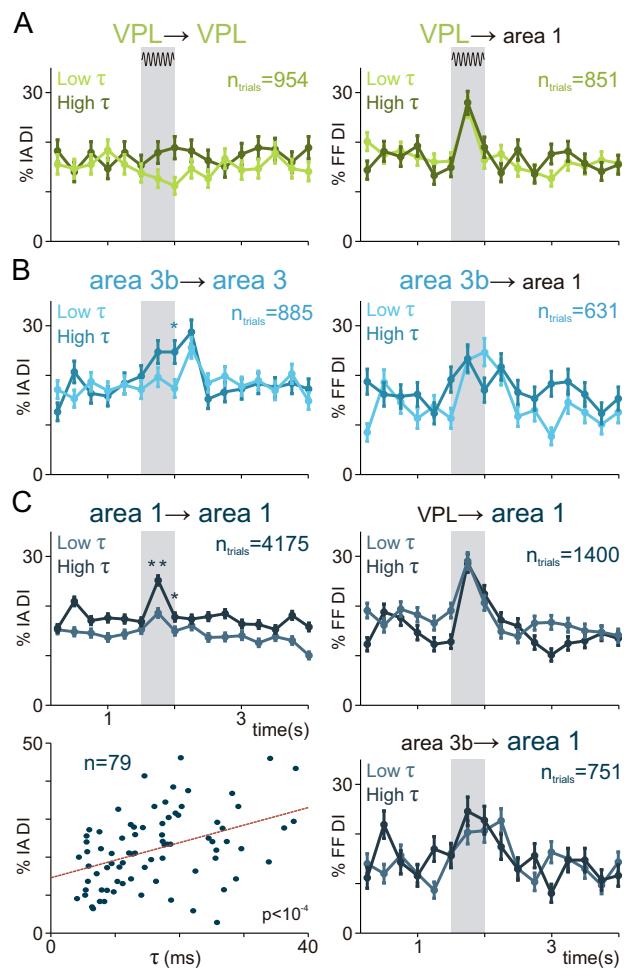

**Figure 6: Cortical intra-area interactions area associated with intrinsic timescales.** Percentage of intra-area incoming unidirectional interactions (left) and feedforward interactions (right) in each area during the task for two groups of neurons (low [light-colored] and high [dark] intrinsic timescale neurons). Error bars denote the SEM and shaded areas highlight the stimulus period (0-0.5 s). (A) Effect of VPL intrinsic timescales. Left: Right: Percentage of intra-VPL unidirectional DI. Right: Percentage of feedforward information from VPL to area 1. (B) Effect of area 3b intrinsic timescales. Left: Percentage of intra-area 3b unidirectional incoming interactions. Right: Percentage of feedforward interactions from area 3b to area 1. (C) Effect of area 1 intrinsic timescales. Left top: Percentage of intra-area 1 unidirectional incoming information. Left bottom: Scatter plot highlighting the correlation ($\rho=0.43$, n=79, $P<10^{-4}$) between the intrinsic timescale and the percentage of incoming intra-area interactions across area 1 neurons during the first half of the stimulus period. Right top: Percentage of feedforward information from VPL to area 1. Right bottom: Percentage of feedforward information from area 3b to area 1.

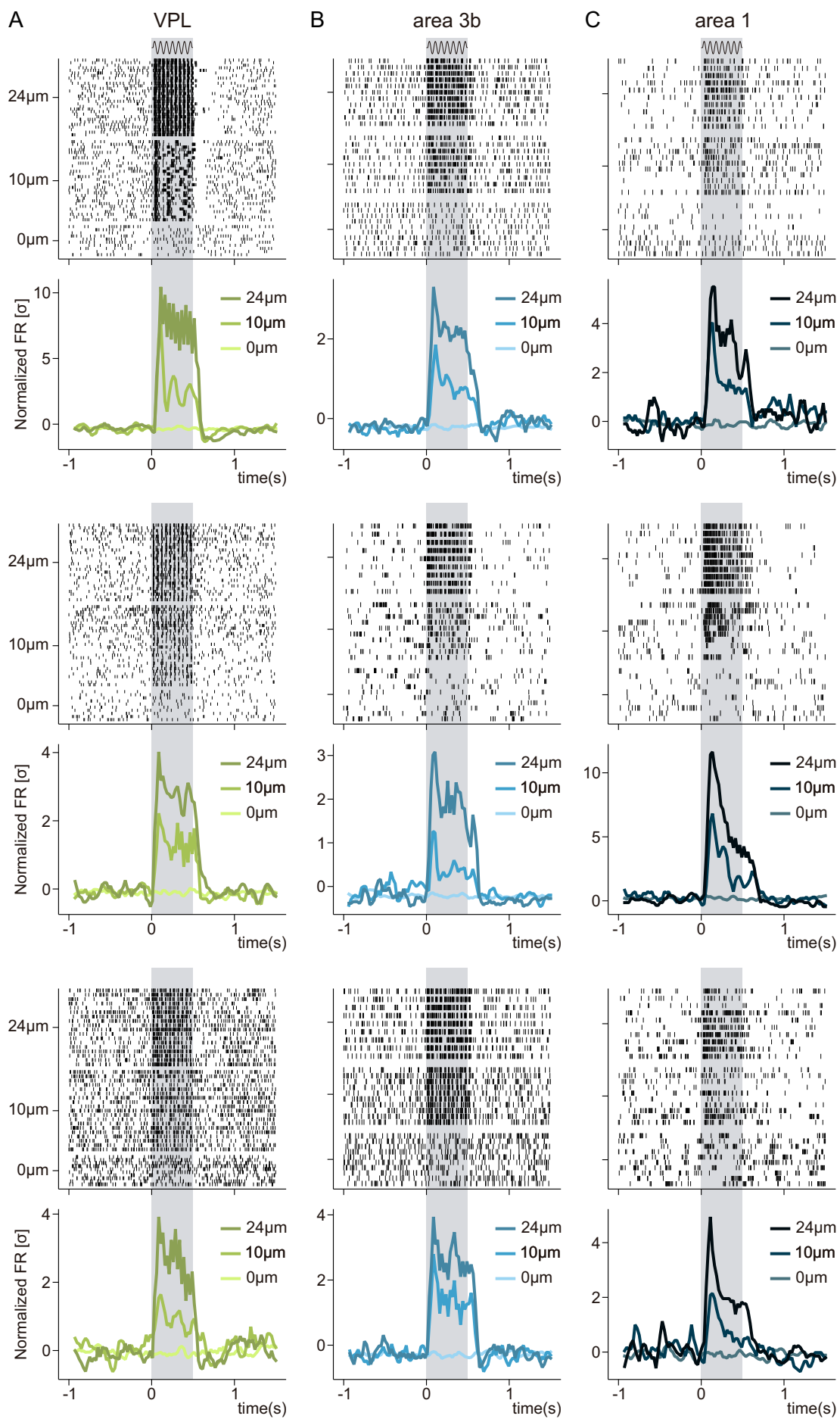

**Figure S1**: **Exemplary raster plots from neurons recorded in VPL, area 3b and area 1** (A-C) Raster plots of nine neurons recorded from VPL (A), area 3b (B) and area 1 (C). Normalized neuronal activity is shown below each raster. In the raster panel, black ticks represent neuronal spikes while colored rectangles represent the stimulation period. On the other hand, in the activity panel, colored lines represent the average of normalized neuronal activity obtained from trials with the same stimulus amplitude and gray rectangles represent the stimulation period.

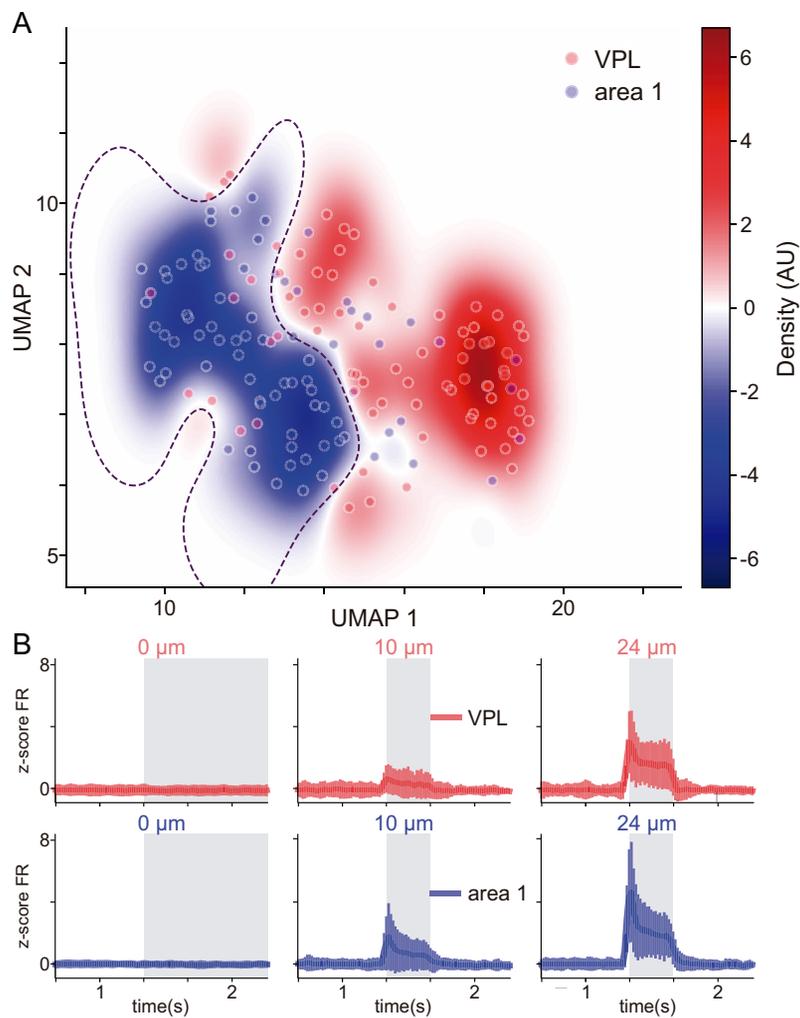

**Figure S2: Classification of the neural activity of VPL and area 1 by a SVM.** A) Density map of the 2d-reduced firing rate by UMAP for VPL (red) and area 1 (blue). The dashed line shows the decision boundary determined by a nonlinear support vector machine. Decision boundary was determined by the 75% of dataset as training. B) Normalized firing rate of the recorded neurons for VPL (top) and area 1 (bottom) for the amplitudes 0 (left) 10 μm (middle) and 24 μm (right) used as input for UMAP. The vertical lines indicate the standard deviation while the center line the population mean.

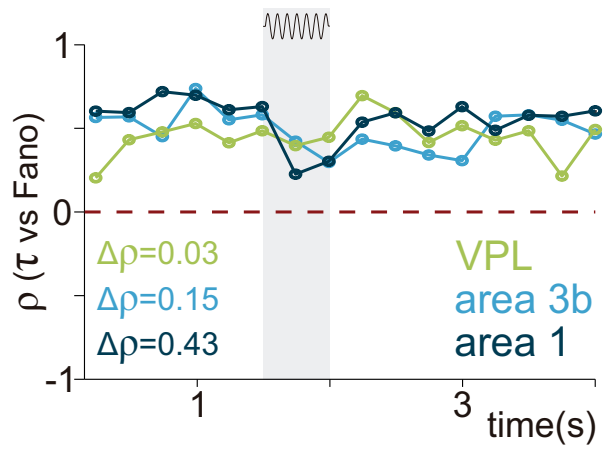

**Figure S3: Timescale and variability correlation in VPL and Area 1.** Spearman's correlation ($\rho$) between the intrinsic timescale and the time-varying Fano factor during the task. Green, VPL; light blue, area 3b; dark blue, area 1. During the first half of the stimulus period (1.5-2s), the correlation between the intrinsic timescale and the Fano factor in area 1 exhibited a larger drop ($|\Delta\rho|=0.42$) with respect to pre-stimulus baseline as compared to area 3b ($|\Delta\rho|=0.15$) and VPL ($|\Delta\rho|=0.03$).

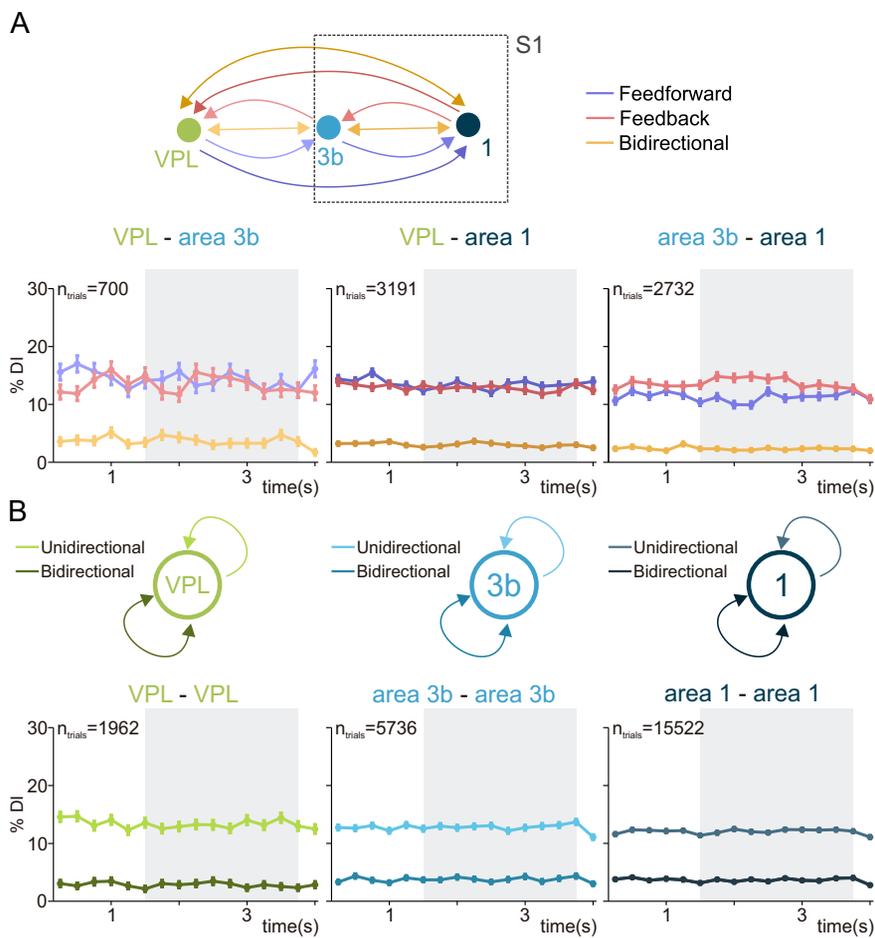

**Figure S4: Inter and intra-area directional interactions in the thalamocortical network during correct rejection (CR) trials.** (A) Scheme showing feedforward, feedback and zero-lag interactions across the areas VPL, 3b and 1 of the somatosensory network. Darker colors same as Fig. 4. (B) Percentage of feedforward (blue) and feedback (red) DI, and zero-lag synchronization (orange) are shown along the task timeline during correct rejections (CR) trials. Left: VPL-area 3b pairs (n=700 trials). Middle: VPL-area 1 pairs (n=3191 trials). Right: area 3b-area 1 pairs (n=2732 trials). Error bars denote the SEM. (C) Scheme showing unidirectional and zero-lag interactions within the areas VPL, areas 3b and 1. (D) Percentage of DI trials for unidirectional (lighter color) and bidirectional zero-lag interactions (darker color) are shown throughout the task during CR trials. Left: VPL-VPL pairs (n=1962 trials). Middle: area 3b-area 3b pairs (n=5736 trials). Right: area 1-area 1 pairs (n=15522 trials).

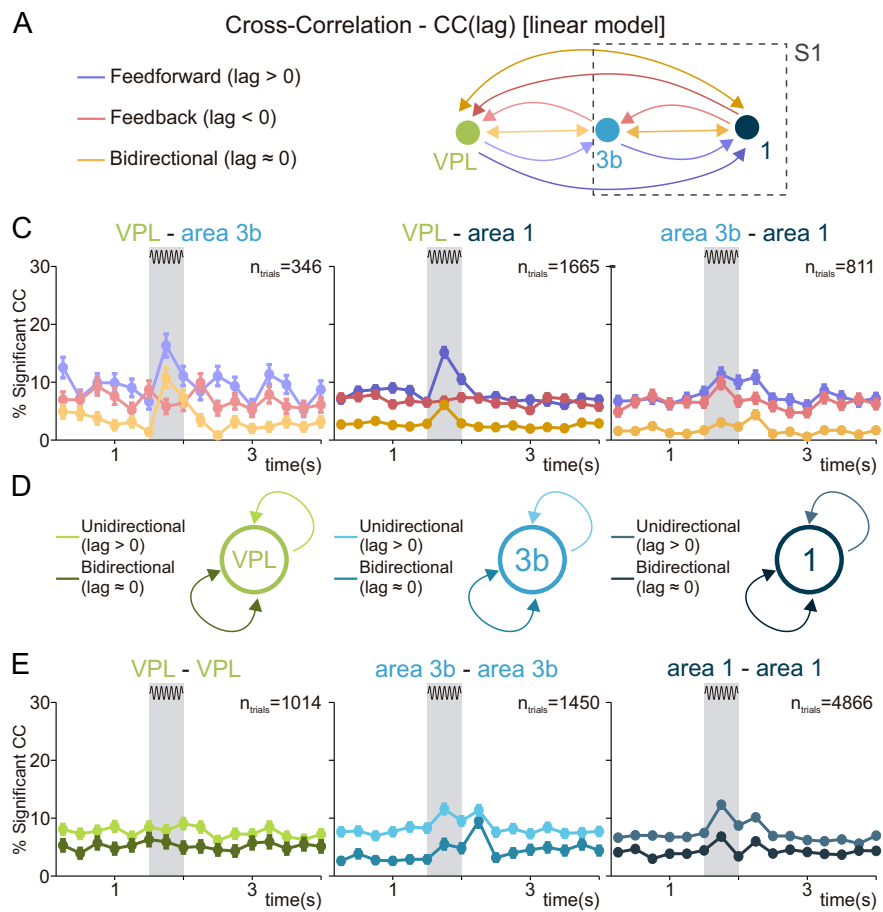

**Figure S5: Inter and intra-area cross-correlations in the thalamocortical network during simultaneously recorded trials of area 3b and 1.** (A) Scheme showing feedforward, feedback and bidirectional interactions across VPL, area 3b and area 1 defined via the linear connectivity measure of cross-correlation (CC) instead of directed information but following the very same non-parametric significance analysis (See Fig. 4A). The arrows connecting VPL and area 1 are slightly darker colors to indicate the relationships in (B). Darker colors represent direct interactions without passing through area 3b as an intermediary. (B) Percentage of significant CC trials for feedforward (blue), feedback (red) and bidirectional (orange) interactions are shown along the task during supra-threshold hits. Left: VPL-area 3b neuron pairs (n=346 trials). Middle: VPL-area 1 pairs (n=1665 trials). Right: area 3b-area 1 pairs (n=811 trials). Error bars denote the SEM. (C) Scheme showing unidirectional and bidirectional interactions within VPL (left), area 3b (middle) and area 1 (right). (D) Percentage of significant CC trials for unidirectional (lighter color) and bidirectional interactions (darker color) are shown throughout the task during supra-threshold hit trials. Left: VPL-VPL pairs (n=1014 trials). Middle: area 3b-area 3b pairs (n=1450 trials). Right: area 1-area 1 pairs (n=4866 trials). Error bars denote the SEM.

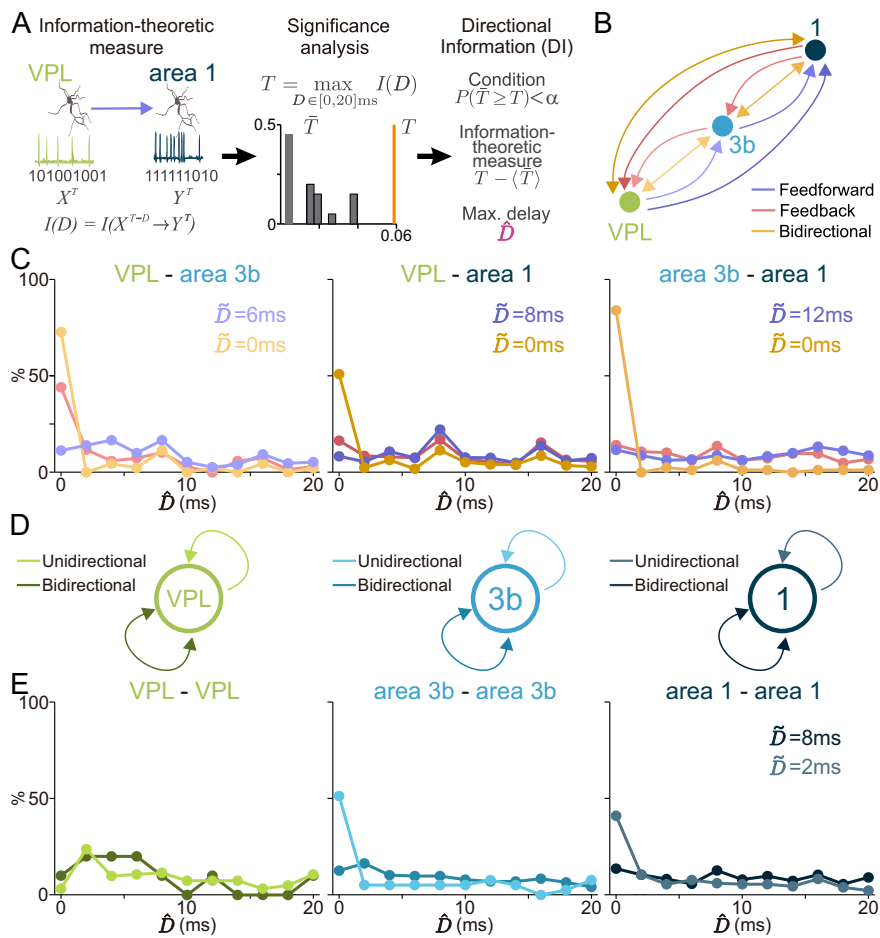

**Figure S6: Inter and intra-area directional interactions delays in the thalamocortical network.** (A), (B) and (D) same as Fig. 4 in structure and color scheme. (C) and (E) Histograms of DI delays and median values for the DI types during the interval 0-0.25s of the stimulus period. Histograms are represented in percentage for all tested delays (0:2:20 ms). Median values are only displayed in cases where the percentage of DI showed a significant increase with respect to the pre-stimulus period (Fig. 4). (C) Delays for inter-area interactions. Left: Delay histogram for VPL-area 3b (n=346 delay values). Middle: VPL-area 1 (n=1665 delay values). Right: area 3b-area 1 (n=811 delay values). (E) Delays for intra-area interactions. Left: Delay histogram for VPL-VPL (n=1014 delay values). Middle: Delay histogram for area 3b-area 3b (n=1450 delay values). Right: area 1- area 1 (n=4866 delay values).